\documentclass[aps,prd,10pt,notitlepage,nofootinbib,superscriptaddress,showkeys,showpacs]{revtex4-1}

\usepackage{amsmath,amssymb}
\usepackage[english]{babel}
\usepackage{graphicx,color}
\usepackage{xspace}
\usepackage{graphicx}
\usepackage{pifont,dsfont}
\usepackage{marvosym}
\usepackage{slashed}
\usepackage{subfigure}

\newcommand{\be}{\begin{equation}}
\newcommand{\ee}{\end{equation}}
\newcommand{\bqa}{\begin{eqnarray}}
\newcommand{\eqa}{\end{eqnarray}}
\newcommand{\bea}{\begin{eqnarray}}
\newcommand{\eea}{\end{eqnarray}}

\DeclareMathOperator{\tr}{tr}

\DeclareMathOperator{\Wg}{Wg}

\begin{document}

\title{\Large \bf Tensor models from the viewpoint of matrix models: the case of the Gaussian distribution}

\author{{\bf Valentin Bonzom}}\email{bonzom@lipn.univ-paris13.fr}
\affiliation{LIPN, UMR CNRS 7030, Institut Galil\'ee, Universit\'e Paris 13, Sorbonne Paris Cit\'e,
99, avenue Jean-Baptiste Cl\'ement, 93430 Villetaneuse, France, EU}
\author{{\bf Fr\'ed\'eric Combes}}\email{frederic.combes@ens-lyon.fr}
\affiliation{Perimeter Institute for Theoretical Physics, 31 Caroline St. N, ON N2L 2Y5, Waterloo, Canada}

\date{\small\today}

\begin{abstract}
\noindent Observables in random tensor theory are polynomials in the entries of tensor of rank $d$ which are invariant under $U(N)^d$. It is notoriously difficult to evaluate the expectations of such polynomials, even in the Gaussian distribution. In this article, we introduce singular value decompositions to evaluate the expectations of polynomial observables of Gaussian random tensors. Performing the matrix integrals over the unitary group leads to a notion of effective observables which expand onto regular, matrix trace invariants. Examples are given to illustrate that both asymptotic and exact new calculations of expectations can be performed this way.

\end{abstract}

\medskip


\maketitle

\section*{Introduction}

A random tensor is a $d$-dimensional array of random variables. The joint probability distribution is typically chosen to be $U(N)^d$-invariant. Just like unitary invariance in random (Hermitian or complex) matrices selects matrix traces as natural invariant observables, the invariance under $U(N)^d$ points to a family of invariant observables, which are polynomials in the tensor entries. Those observables are labeled by connected, $d$-regular, edge-colored graphs, and one is interested in the large $N$ expansion of their expectations.

In matrix models \cite{mm-review-difrancesco}, expectations of observables $\tr M^n$ can be expanded onto Feynman graphs, actually maps whose genus is the exponent of $N$ in the expansion at large $N$. The Feynman expansion also applies to random tensors and gives rise to $(d+1)$-regular edge-colored graphs \cite{Gur3, Gur4, uncoloring}. Each Feynman graph inherits an $N$-dependent weight, the exponent of $N$ being called the \emph{degree} (which reduces to the genus at $d=2$). Those graphs are suited for a combinatorial analysis: the classification of colored graphs of fixed degree has been obtained in \cite{GurauSchaeffer} and the singularities of the generating functions are now known.

However, the calculation of expectations of invariant polynomials is still a challenging issue. Even in the simplest case, i.e. the Gaussian distribution, and even if one only asks for the leading order at large $N$ of an expectation, it is necessary to $i)$ identify the exponent of $N$ for the dominant contributions, $ii)$ find the numerical coefficient which counts the number of dominant contributions (dominant Wick contractions). Both problems are bubble-dependent.

A few cases have been studied. The so-called melonic polynomials all have a single leading order Wick contraction (as well as a generalization of this family including nearly-melonic polynomials) \cite{uncoloring}. A family of polynomials has been found in \cite{MeandersTensors} whose numbers of leading order Wick contractions are numbers of meandric systems.

It is worth emphasizing the importance of the Gaussian distribution in random tensor models. It was proved in \cite{universality} that any $U(N)^d$-invariant distribution satisfying some assumptions on its dependence with $N$ becomes Gaussian at large $N$ (the covariance does depend on the initial distribution) -- models which do not satisfy those assumptions (and not Gaussian at large $N$) are discussed in \cite{new1/N}. Moreover, models which do not have a trivial bare covariance, and therefore have a renormalization group flow, tend to be asymptotically free: they have a Gaussian fixed point (this happens surprisingly more often than in ordinary quantum field theory) \cite{FRGforTGFT, RenormalizableGeloun, BetaFunctionD=4, EpsilonCarrozza}. It is appealing to think this somehow comes from the universality theorem of \cite{universality}.

In addition to the purely combinatorial analysis of \cite{GurauSchaeffer}, matrix model techniques have recently been employed in tensor models. They apply to models with quartic interactions for which the Hubbard-Stratanovich transformation leads to multi-matrix models \cite{DSQuartic, BeyondPert, GenericQuartic}. Furthermore, as argued in \cite{DSSD}, the behavior of generic models can then be deduced from those of the quartic ones (universality phenomenon).

The relationship of such multi-matrix models to ordinary matrix models has been studied in \cite{FullyPacked} where they are described in terms of fully packed loop configurations on random surfaces generated by $U(\tau)$ models. The results of random tensor theory were applied in this context, leading to a classification of loop configurations which corresponds to the classification of colored graphs. This basically relied on interpreting a collection of matrices as a tensor.

Here we focus on the Gaussian distribution instead, and use another direct connection between random tensors and matrices. Splitting the set of $d$ indices of the tensor into two sets, we can consider the tensor as a (typically rectangular) matrix of size $N^{p}\times N^{d-p}$ and then use the singular value decomposition. However, tensor models are not $U(N^p)$ or $U(N^{d-p})$ invariant, meaning that the integrals over the angular degrees of freedom are non-trivial. When they can be performed at fixed singular values, an effective theory on the singular values is obtained.

Although it is a simple idea, it is only the first time that it is applied explicitly to tensor models. As a first step in this program, we consider here the case of the calculation of expectations of polynomial observables in a Gaussian distribution. The angular integrals are integrals of polynomials in the unitary matrix elements and the formula of \cite{Collins} can be applied. In concrete situations where the degree of the polynomial is not too large, exact results are obtained. We also show that large $N$ behaviors can be extracted using the diagrammatic method of \cite{DiagrammaticU(N)} on a family of observables which generalize the melonic ones (basically, trees, like in the melonic case, formed by the gluing of matrix trace-invariants).

In addition to providing so far unexplored relationships between random tensors and matrices, the approach of \cite{FullyPacked} and the present paper shows the difficulties faced in random tensor theory in the familiar context of matrix models.

\section{Gaussian expectations in random tensor theory} \label{sec:Gaussian}

The work presented in \cite{FullyPacked} relied on the simple observation that a set of matrices with a unitary symmetry among them can be seen as a tensor equipped with independent unitary transformations on its indices. There is another simple way to relate tensor to matrices, which is to consider a tensor as a (typically rectangular) matrix between a subset of indices to the subset of the other indices. For a tensor on $d$ indices, seen as a linear form on $\bigoplus_{i=1}^d V_{i}$, one picks up a subset $\mathcal{C}=\{i_1,\dotsc,i_{|\mathcal{C}|}\} \subset \{1,\dotsc,d\}$ and denote its complement $\{k_1,\dotsc,k_{d-|\mathcal{C}|}\} = \{1,\dotsc,d\}\setminus \mathcal{C}$. One defines $M$ as a linear application from $\bigoplus_{p=1}^{d-|\mathcal{C}|} V_{k_p}$ to $\bigoplus_{j=1}^{|\mathcal{C}|} V_{i_j}$. 

\begin{figure}
\includegraphics[scale=.5]{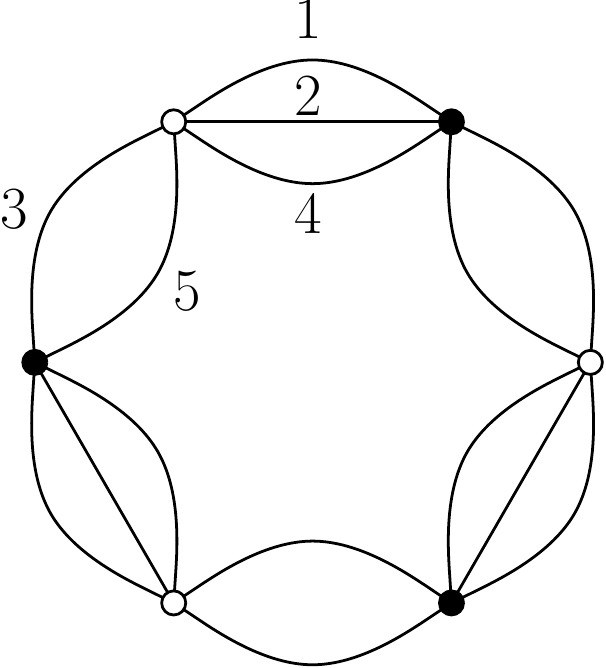}
\caption{\label{fig:Necklace} This is the necklace on five colors with $\mathcal{C}=\{3,5\}$ and 6 vertices.}
\end{figure}

Of course, there are multiple ways to choose $\mathcal{C}$, but depending on the observables one is interested in, some choices are better than others. We recall that a basis of observables is formed by the polynomials $B(T,\overline{T})$ labeled by bubbles. A \emph{bubble} is a connected bipartite $d$-regular edge-colored graph, with colors $1,\dotsc,d$ such that all $d$ colors are incident to each vertex (see Figure \ref{fig:Necklace}). The polynomial associated to a bubble is built by writing a tensor $T$ for each white vertex and its conjugated tensor $\overline{T}$ for each black vertex. When an edge of color $i$ connects a white to a black vertex, the indices in position $i$ of the corresponding $T$ and $\overline{T}$ are identified (with a Kronecker delta) and summed over from $1$ to $N$. Such polynomials are invariant under $U(N)^d$, \cite{uncoloring}.

Loosely speaking, a ``good'' choice of the subset $\mathcal{C}$ relative to $B(T,\overline{T})$ is one that minimizes the degrees of freedom entering $B$. A natural way of separating the degrees of freedom of $T,\overline{T}$ is to use the singular value decomposition with respect to $\mathcal{C}$, i.e. $M=UDV$, where $U,V$ are unitary and $D$ is made of a square diagonal matrix completed with some rows or columns of zeros. For a few families of polynomials $B$, the choice of $\mathcal{C}$ is obvious, e.g. when there exists one such that $B(T,\overline{T}) = B(D)$ is a function of the singular values only.

This last situation however requires to focus on polynomials of the form $\tr (MM^\dagger)^n$ which we call \emph{necklaces}. Their bubbles consist of single cycles on $2n$ vertices connected by two types of multiple edges, those with colors in $\mathcal{C}$ and those with colors in the complementary subset (see example in figure \ref{fig:Necklace}).

Generically, a polynomial $B(T,\overline{T}) = B(U,V,D)$ depends on all the degrees of freedom of the singular value decomposition. This is simply because a polynomial invariant under the action of $U(N_1)\times \dotsb \times U(N_d)$ is generically not invariant under $U(\prod_{j=1}^{|\mathcal{C}|} N_{i_j}) \times U(\prod_{p=1}^{d-|\mathcal{C}|} N_{k_p})$. Notice that the Gaussian measure however is independent of the angular matrices $U,V$, i.e. $T\cdot \overline{T} = \tr MM^\dagger$ for any choice of $\mathcal{C}$. Therefore the \emph{Gaussian} expectation of a polynomial reads
\begin{equation}
\langle B(T,\overline{T})\rangle = \frac1Z\, \int d\mu(\{\lambda_i\})\,e^{-N^{d-1}\sum_i \lambda_i^2} \int dU\,dV\ B(U,V,\{\lambda_i\}).
\end{equation}
Here $\{\lambda_i\geq0\}$ denotes the set of singular values, $d\mu(\{\lambda_i\})$ the measure inherited from the change of variables, while $dU, dV$ are the Haar measures on unitary matrices of the appropriate sizes.

This provides a notion of effective polynomial of $B$ with respect to $\mathcal{C}$ which is a function over the singular values,
\begin{equation} \label{AngularIntegral}
B_{\mathcal{C}}(\{\lambda_i\}) = \int dU\,dV\ B(U,V,\{\lambda_i\}).
\end{equation}
Then, the expectation of $B$ is just the expectation of $B_{\mathcal{C}}$ in the complex Gaussian Wishart ensemble (also known as the Laguerre ensemble, from the family of relevant orthogonal polynomials for this measure). The integral over the angular variables results in an expansion onto $U(\prod_{j=1}^{|\mathcal{C}|} N_{i_j}) \times U(\prod_{p=1}^{d-|\mathcal{C}|} N_{k_p})$-invariants,
\begin{equation}
B_{\mathcal{C}}(\{\lambda_i\}) = \sum_k \sum_{l_1,\dotsc,l_k} c^{(B)}_{l_1,\dotsc,l_k}(N_1,\dotsc,N_d)\ \Bigl(\sum_{i_1} \lambda_{i_1}^{2l_1}\Bigr) \dotsm \Bigl(\sum_{i_k} \lambda_{i_k}^{2l_k}\Bigr),
\end{equation}
and therefore
\begin{equation}
\langle B(T,\overline{T}) \rangle = \sum_k \sum_{l_1,\dotsc,l_k} c^{(B)}_{l_1,\dotsc,l_k}(N_1,\dotsc,N_d)\ \langle \Bigl(\sum_{i_1} \lambda_{i_1}^{2l_1}\Bigr) \dotsm \Bigl(\sum_{i_k} \lambda_{i_k}^{2l_k}\Bigr) \rangle_{\text{Laguerre}}.
\end{equation}

Taking all the tensor indices to have range $N$, the expectation of the product of single-trace invariants factorizes as the product of the expectations in the large $N$ limit. There are two possible cases,
\begin{equation}
\langle \sum_i \lambda_i^{2l} \rangle = \langle \tr (MM^\dagger)^{l} \rangle \sim_{N\to\infty} \begin{cases} 1 & \text{if $|\mathcal{C}| \neq d-|\mathcal{C}|$},\\ \operatorname{Cat}_l & \text{if $|\mathcal{C}| = d-|\mathcal{C}|$}. \end{cases}
\end{equation}
The symbol $\sim$ here also means up to some power of $N$ and $\operatorname{Cat}_l$ is the $l$-th Catalan number. When the matrix is rectangular, it is quite unbalanced since the ratio of its dimensions goes to either zero or infinity, and as a result a single Wick contraction survives at large $N$. When $\mathcal{C}$ (or its complement) is a singlet, it means that $B$ is a cycle of melons inserted on a fixed color and the result follows from \cite{universality}. In the other cases, it is an application of the results of \cite{new1/N} to the Gaussian distribution (which rely on the fact that there is a melonic subgraph which visits all the vertices). Only when $M$ is a square matrix one recovers the familiar Catalan numbers of Gaussian matrix models.


Several methods have been developed to deal with integrals over the unitary group, \cite{U(N)IntegralsAubert, ZinnZuber}. For our purpose, since the function $B(U,V,\{\lambda_i\})$ is polynomial in the matrix elements of $U,U^\dagger,V,V^\dagger$, the following formula \cite{Collins} seems to be the most natural to perform the integral \eqref{AngularIntegral},
\begin{equation} \label{U(N)Integral}
\int_{U(N)} dU\ U_{a_1 \alpha_1} \dotsm U_{a_n \alpha_n}\, \overline{U}_{b_1 \beta_1} \dotsm \overline{U}_{b_n \beta_n} = \sum_{\sigma,\tau\in\mathfrak{S}_n} \delta_{a_1,b_{\sigma(1)}}\dotsm \delta_{a_n, b_{\sigma(n)}}\,\delta_{\alpha_1, \beta_{\tau(1)}} \dotsm \delta_{\alpha_n, \beta_{\tau(n)}}\ \operatorname{Wg}_N(\sigma\tau^{-1}).
\end{equation}
$\sigma$ and $\tau$ run over the symmetric group on $n$ elements $\mathfrak{S}_n$, and $\operatorname{Wg}_N$ is a Weingarten function over $\mathfrak{S}_n$.

In principle, this provides a way to compute the effective polynomial $B_{\mathcal{C}}$, and there even exists a diagrammatic expansion which enables to control the scaling with $N$ of the various terms involved in the sums over permutations \cite{DiagrammaticU(N)}. Yet in practice, the number $n$ of matrix elements of $U$ (and similarly for $V$) cannot be too large as the sum over permutations becomes unmanageable for relatively large $n$.

We now set $d=4$ and $\mathcal{C} = \{2,4\}$ and illustrate the method on two examples: a simple, but exact calculation, and a large $N$ behavior on a family of observables. We write $T_{a_1 a_2 a_3 a_4} = M_{\begin{smallmatrix} a_1\\ a_3\end{smallmatrix} \begin{smallmatrix} a_2\\ a_4\end{smallmatrix}}$ where a column of tensor indices represents a index of $M$ with range $N^2$. Therefore $MM^\dagger = UD^2 U^\dagger$ is a $N^2\times N^2$ matrix $V_1\otimes V_3 \to V_1\otimes V_3$.

\subsection{A simple, exact calculation} \label{sec:ExactCalculation}

If $B$ is a bubble in which the edges of color 2 and 4 always connect the same vertices, then the associated polynomial is a function of $MM^\dagger$. Moreover, the edges of color 2 count the degree in $MM^\dagger$. We consider the following bubble, with $k$ edges of color 2 on the top and $l$ at the bottom of the drawing,
\begin{equation} \label{BubbleEdgeTree}
\begin{array}{c} \includegraphics[scale=.4]{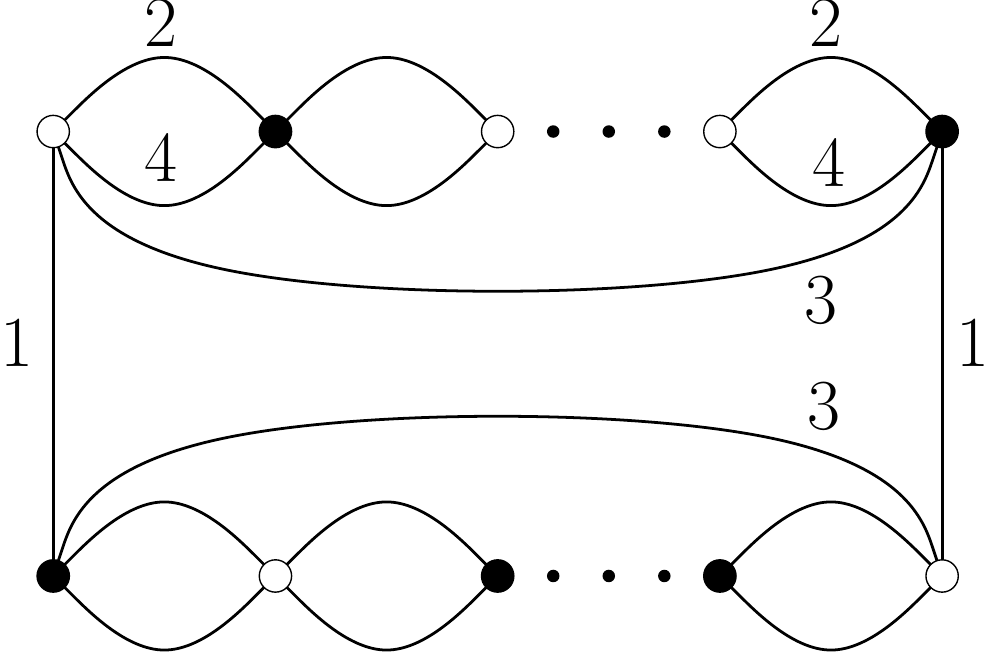} \end{array} = 
\tr_1 \Bigl(\tr_3(MM^\dagger)^{k}\,\tr_3(MM^\dagger)^{l} \Bigr) = \sum_{a_1,a_3,b_1,b_3=1}^N (MM^\dagger)^{k}_{\begin{smallmatrix} a_1\\ a_3\end{smallmatrix} \begin{smallmatrix} b_1\\ a_3\end{smallmatrix}}\ (MM^\dagger)^{l}_{\begin{smallmatrix} b_1\\ b_3\end{smallmatrix} \begin{smallmatrix} a_1\\ b_3\end{smallmatrix}},
\end{equation}
using an obvious partial trace notation. By writing $MM^\dagger = UD^2U^\dagger$ with $D^2 = \operatorname{diag}(\{\lambda_{\begin{smallmatrix} \alpha_2\\ \alpha_4\end{smallmatrix}}\})$, we find
\begin{equation}
\sum U_{\begin{smallmatrix} a_1\\ a_3\end{smallmatrix} \begin{smallmatrix} \alpha_2\\ \alpha_4\end{smallmatrix}} \lambda_{\begin{smallmatrix} \alpha_2\\ \alpha_4\end{smallmatrix}}^{2k} U^\dagger_{\begin{smallmatrix} \alpha_2\\ \alpha_4\end{smallmatrix} \begin{smallmatrix} b_1\\ a_3\end{smallmatrix}}\, U_{\begin{smallmatrix} b_1\\ b_3\end{smallmatrix} \begin{smallmatrix} \beta_2\\ \beta_4\end{smallmatrix}} \lambda_{\begin{smallmatrix} \beta_2\\ \beta_4\end{smallmatrix}}^{2l} U^\dagger_{\begin{smallmatrix} \beta_2\\ \beta_4\end{smallmatrix} \begin{smallmatrix} a_1\\ b_3\end{smallmatrix}}
\end{equation}
The integral over $U(N^2)$ is simple enough since there are only two permutations on two elements,
\begin{multline}
\int_{U(N^2)} dU\ U_{\begin{smallmatrix} a_1\\ a_3\end{smallmatrix} \begin{smallmatrix} \alpha_2\\ \alpha_4\end{smallmatrix}}\,U_{\begin{smallmatrix} b_1\\ b_3\end{smallmatrix} \begin{smallmatrix} \beta_2\\ \beta_4\end{smallmatrix}}\,\overline{U}_{\begin{smallmatrix} b_1\\ a_3\end{smallmatrix} \begin{smallmatrix} \alpha_2\\ \alpha_4\end{smallmatrix}}\,\overline{U}_{\begin{smallmatrix} a_1\\ b_3\end{smallmatrix} \begin{smallmatrix} \beta_2\\ \beta_4\end{smallmatrix}} = \Wg_{N^2}(1^2) \Bigl(\delta_{a_1,b_1} + \delta_{a_3,b_3}\delta_{\alpha_2,\beta_2}\delta_{\alpha_4,\beta_4} \Bigr) \\
+ \Wg_{N^2}(2) \Bigl(\delta_{a_3,b_3} + \delta_{a_1, b_1}\delta_{\alpha_2,\beta_2}\delta_{\alpha_4,\beta_4} \Bigr).
\end{multline}
Only the cycle structure of the arguments of the Weingarten functions is retained (they are class functions), so that $1^2$ is the (class of the) identity and $2$ the (class of the) transposition. Moreover,
\begin{equation}
\Wg_{N^2}(1^2) = \frac{1}{N^4-1},\qquad \Wg_{N^2}(2) = -\frac{1}{N^2(N^4-1)}.
\end{equation}
Performing the sums, it comes
\begin{equation}
\sum_{a_1,b_1,a_3,b_3} \int_{U(N^2)} dU\ U_{\begin{smallmatrix} a_1\\ a_3\end{smallmatrix} \begin{smallmatrix} \alpha_2\\ \alpha_4\end{smallmatrix}}\,U_{\begin{smallmatrix} b_1\\ b_3\end{smallmatrix} \begin{smallmatrix} \beta_2\\ \beta_4\end{smallmatrix}}\,\overline{U}_{\begin{smallmatrix} b_1\\ a_3\end{smallmatrix} \begin{smallmatrix} \alpha_2\\ \alpha_4\end{smallmatrix}}\,\overline{U}_{\begin{smallmatrix} a_1\\ b_3\end{smallmatrix} \begin{smallmatrix} \beta_2\\ \beta_4\end{smallmatrix}} = \frac{N}{N^2+1} \bigl(1+\delta_{\alpha_2,\beta_2}\,\delta_{\alpha_4,\beta_4}\bigr),
\end{equation}
which is the tensor that has to be contracted with the singular values. Eventually we arrive at
\begin{equation}
\begin{aligned}
B_{\mathcal{C}} (\{ \lambda_{\begin{smallmatrix} \alpha_2\\ \alpha_4\end{smallmatrix}} \}) &= \frac{N}{N^2+1} \Bigl( \bigl(\sum_{\alpha_2,\alpha_4} \lambda_{\begin{smallmatrix} \alpha_2\\ \alpha_4\end{smallmatrix}}^{2k} \bigr)\,\bigl(\sum_{\beta_2,\beta_4} \lambda_{\begin{smallmatrix} \beta_2\\ \beta_4\end{smallmatrix}}^{2l} \bigr) + \sum_{\alpha_2,\alpha_4} \lambda_{\begin{smallmatrix} \alpha_2\\ \alpha_4\end{smallmatrix}}^{2k+2l} \Bigr)\\
&= \frac{N}{N^2+1} \Bigl(\tr(MM^\dagger)^{k}\,\tr(MM^\dagger)^{l} + \tr (MM^\dagger)^{k+l}\Bigr).
\end{aligned}
\end{equation}

The large $N$ limit of the expectation can be easily extracted. Normalizing the Gaussian with a covariance $1/N^2$, the double-trace term dominates (each trace bringing up a factor $N^2$) and factorizes, so that
\begin{equation}
\langle B(T,\overline{T})\rangle \underset{\text{large $N$}}{=} N^3\ \operatorname{Cat}_{k}\,\operatorname{Cat}_{l}.
\end{equation}
(The scaling factor $N^3$ is due to the fact that the bubble has a single cycle with colors $(1,2)$ and two cycles of colors $(3,4)$, see \cite{new1/N}.)

\subsection{Large $N$ behavior} \label{sec:CatalanTrees}

When the degree of the polynomial in $U$ (and/or $V$) which has to be integrated is large, an exact result becomes difficult to extract and largely depend on the combinatorics of the initial bubble. Even the asymptotics seems difficult to evaluate. However, in some cases, the method can lead to the large $N$ limit.

\begin{figure}
\includegraphics[scale=.35]{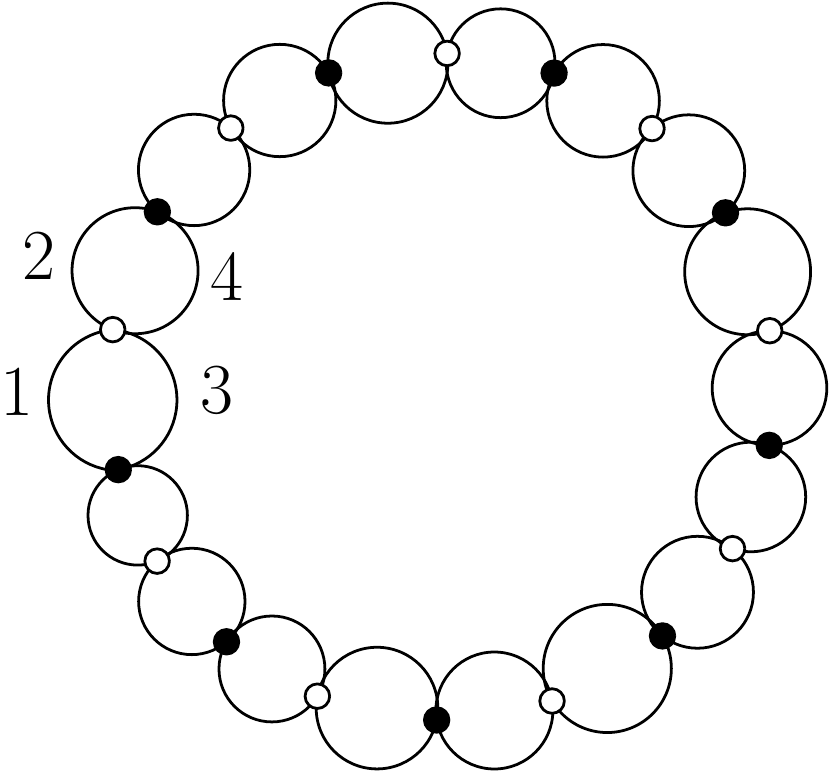}
\hspace{3cm}
\includegraphics[scale=.35]{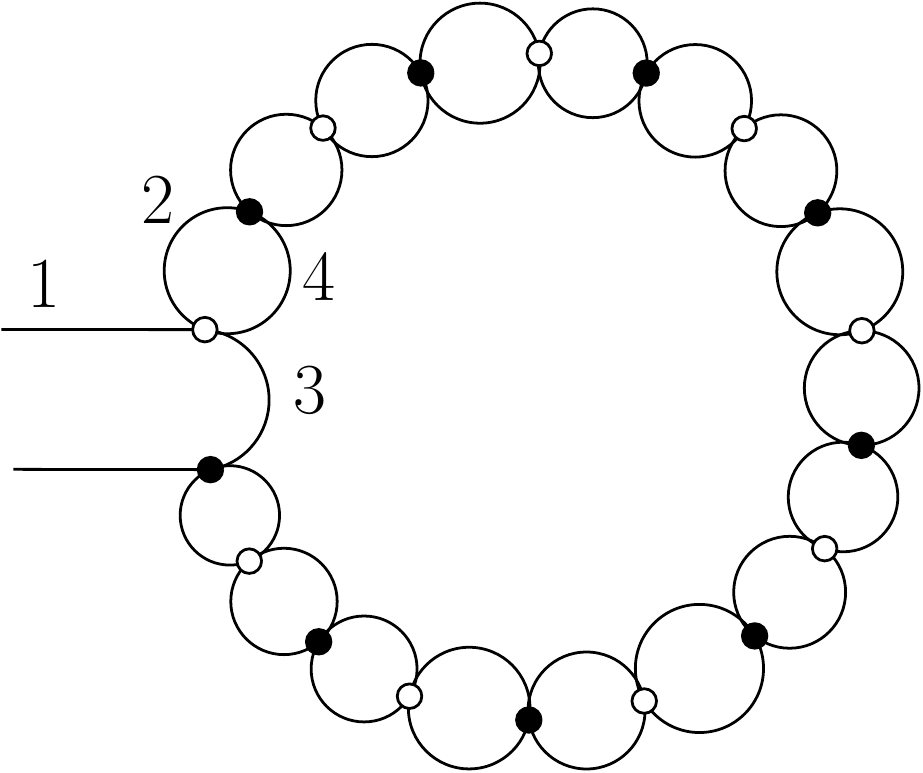}
\caption{\label{fig:OpenNecklace} On the left is a necklace and on the right an open necklace of color 1.}
\end{figure}

We consider a family of observables built in the following way. We first define the \emph{open necklace of color $i$} as the necklace with a line of color $i$ cut into two halves (see figure \ref{fig:OpenNecklace}). Then the construction starts with a necklace of length $k_1$. A random edge of color 1 or 3 is deleted and instead an open necklace of the appropriate color is attached so as to get a bipartite graph. On this new graph, a randomly chosen edge of color 1 or 3 is removed and replaced by an open necklace of the same color. The family of interest in this section consists of graphs built by continuing this process recursively a finite number of times.

To avoid redundancies in the construction, one can consider the rooted observables. The construction starts with an open necklace of color 1. Then the process of piling up open necklaces is similar, except that after an insertion on the color $i$, further insertions on the newly created edge of this color and incident to the black vertex of the inserted open necklace are forbidden. All other edges of color 1 and 3 are called active edges. This way, there is a partial order between the open necklaces used to build up the observables. A necklace (a child) is smaller than another one (a parent) if it is inserted on an active edge of the other one. Every necklace (but the root one) has a single parent.

\begin{figure}
\includegraphics[scale=.35]{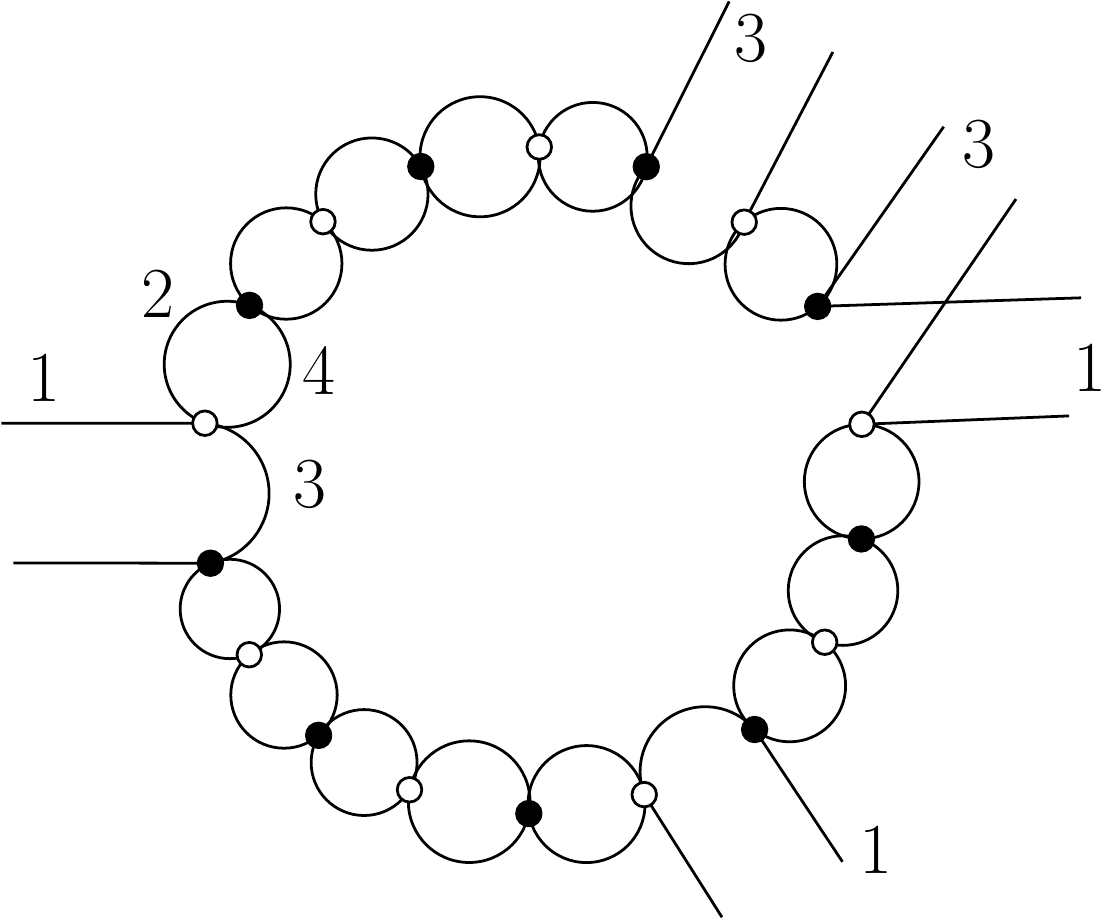}
\caption{\label{fig:TreeVertex} This represents a typical necklace within an observable, which is mapped to a vertex of the corresponding tree. If we assume the edge connected to the parent is the open one incident to the black vertex and of color 1 on the left, then the distances labeling the corners (going counter-clockwise) read 3, 2, 0, 1, 3, 0. The final 0 is only here when there is a necklace insertion on the open edge of color 1 on the left, incident to the white vertex.}
\end{figure}

There is a simple bijection between the set of such rooted observables and a family of rooted, corner-labeled, plane trees. It is a generalization of the bijection between melonic graphs and trees, explained in \cite{critical-colored}, in the sense that the case of melonic graphs with melons on the colors 1 and 3 only will be recovered by removing the labels on the trees. The tree associated to an observable simply represents the relations of partial order between the necklaces. Vertices correspond to open necklaces and edges of color 1 and 3 represent the child/parent relationships. Notice that a typical necklace however has several edges of color 1 and 3 so that further decorations are required to keep track of the specific edges on which insertions are performed. One defines the distance between two edges incident to the same vertex and both connected to children of this vertex as the number of edges of color 2 which separate the two insertions on the necklace corresponding to this vertex. The distance at a vertex $v$ between an edge connected to a children of this vertex and an edge connected to the parent of $v$ is the number of edges of color 2 around the necklace corresponding to $v$ between the edge incident to a black vertex which goes to the parent necklace and the edge on which the insertion of the children necklace is performed. In the tree, those distances are included by preserving the cyclic ordering of the insertions around each vertex, so that they label the corners. An example is given in figure \ref{fig:TreeVertex}.


The bubble considered in the equation \eqref{BubbleEdgeTree} is represented by the tree with two vertices and a single edge of color 1 between them. It has two corners, one with label $k$ and one with label $l$. (The necklace itself could be seen as the tree consisting of a single vertex, which has no corner, although an integer is needed for the length.) If $C(\mathcal{T})$ denotes the set of corners of the tree $\mathcal{T}$, $V(\mathcal{T})$ its set of vertices, and $\{k_c\}_{c\in C(\mathcal{T})}$ the set of integers attached to the corners of $\mathcal{T}$, the total length around the vertex $v\in V(\mathcal{T})$ is $k_v = \sum_{\text{$c$ around $v$}} k_c$. Then
\begin{equation} \label{CatProd}
\langle B_{\mathcal{T}}(T,\overline{T})\rangle \underset{\text{large $N$}}{=} \prod_{v\in V(\mathcal{T})} \operatorname{Cat}_{k_v}.
\end{equation}

To prove this, we will show that the removal of a leaf with label $l$ and its incident edge in the tree amounts to factorizing the Catalan number $\operatorname{Cat}_l$. Zooming on the explicit dependence of $B_{\mathcal{T}}$ at such a leaf, say with incident edge of color 1, we have
\begin{equation}
B_{\mathcal{T}} = \sum_{a_1,b_1=1}^N \Bigl[\tr_3 (MM^\dagger)^l\Bigr]_{a_1 b_1}\,\Bigl[f(MM^\dagger)\Bigr]_{b_1 a_1} = \sum U_{\begin{smallmatrix} a_1\\ a_3\end{smallmatrix} \begin{smallmatrix} \alpha_2\\ \alpha_4\end{smallmatrix}} \lambda_{\begin{smallmatrix} \alpha_2\\ \alpha_4\end{smallmatrix}}^{2l} \overline{U}_{\begin{smallmatrix} b_1\\ a_3\end{smallmatrix}\begin{smallmatrix} \alpha_2\\ \alpha_4\end{smallmatrix}}\,\Bigl[f(UD^2U^\dagger)\Bigr]_{b_1 a_1}.
\end{equation}
Here $f$ is a matrix-valued function corresponding to chopping off the leaf and half the incident edge.

To perform the integral over $U$, we use the diagrammatic method of \cite{DiagrammaticU(N)}. First, one gets a diagram corresponding to the integrand and then the expansion of the integral \eqref{U(N)Integral} is obtained as a sum over decorations of this diagram by additional edges (akin to Wick contractions of Feynman diagrams). The matrix element of $U$ is represented by an edge between a black vertex (corresponding to the left indices, of colors 1, 3) and a white vertex (corresponding to the right indices, of colors 2, 4). For $\overline{U}$, the edge is simply decorated with a star. The matrix elements of powers of $D^2$ are represented as edges with a box on them. Finally the indices of colors 1, 3 are drawn explicitly as half-edges. This way,
\begin{equation}
\sum_{\alpha_2,\alpha_4=1}^N U_{\begin{smallmatrix} a_1\\ a_3\end{smallmatrix} \begin{smallmatrix} \alpha_2\\ \alpha_4\end{smallmatrix}} \lambda_{\begin{smallmatrix} \alpha_2\\ \alpha_4\end{smallmatrix}}^{2l} \overline{U}_{\begin{smallmatrix} b_1\\ b_3\end{smallmatrix}\begin{smallmatrix} \alpha_2\\ \alpha_4\end{smallmatrix}} = \begin{array}{c} \includegraphics[scale=.65]{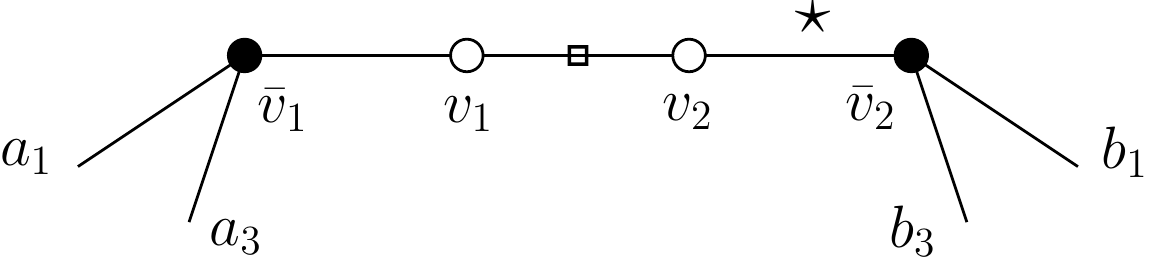} \end{array}
\end{equation}
The integral is performed using \eqref{U(N)Integral} and $n$ denotes the number of matrix elements of $U$. Each term in the resulting sum over permutations $\sigma, \tau\in\mathfrak{S}_n$ is represented as a diagram obtained from the one of the integrand by adding a dashed edge for each Kronecker delta. The permutation $\sigma$ (respectively $\tau$) is thus represented by dashed edges connecting black (respectively white) vertices two by two. Each diagram is then weighted with the Weingarten function $\Wg_{N^2}(\sigma\tau^{-1})$.

\begin{figure}
\includegraphics[scale=.5]{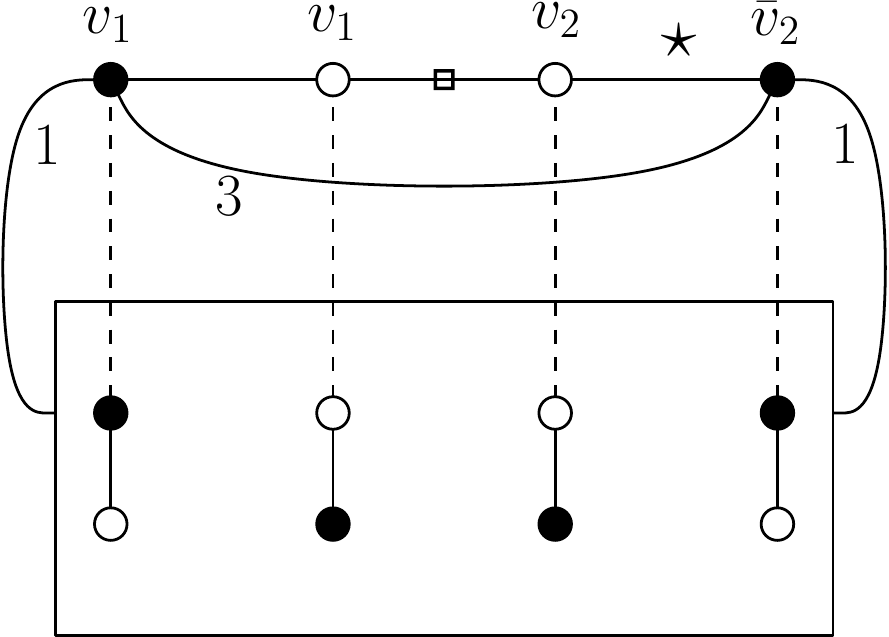}
\hspace{4cm}
\includegraphics[scale=.5]{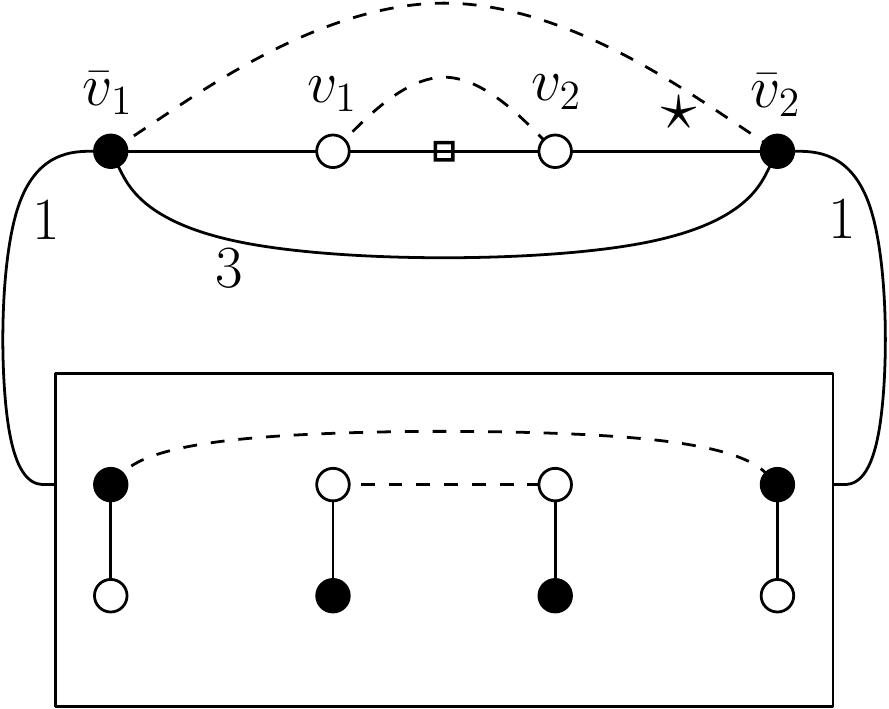}
\caption{\label{fig:LeafContraction} The solid, non-colored edges represent matrix elements of $U$, and $\overline{U}$ with a star, edges with boxes represent singular values. Contractions of indices of colors 1, 3 are represented with edges incident to black vertices and the dashed edges correspond to the Kronecker deltas induced by the permutations $\sigma, \tau$ in \eqref{U(N)Integral}. The vertices $v_1, v_2, \bar{v}_1, \bar{v}_2$ represent the matrix indices on a leaf of a tree and the rectangle contains the rest of the tree. The diagram on the right is obtained from the one of the left by cutting the dashed edges which are drawn and reconnecting them so that $v_1$ is connected to $v_2$ and $\bar{v}_1$ to $\bar{v}_2$.}
\end{figure}

Consider a diagram where $\bar{v}_1$ and $v_1$ are not connected to $\bar{v}_2$ and $v_2$, but rather to other vertices from other parts of the tree, as in the left of figure \ref{fig:LeafContraction}. The rectangle there stands for an arbitrary configuration of the remaining dashed edges. We are going to compare this situation to another diagram obtained from it by cutting the four dashed edges in halves and reconnecting them as on the right of figure \ref{fig:LeafContraction} (all dashed lines which are not drawn are untouched). We want to compare their amplitude as a function of $N$. The $N$-dependence is found by counting the number of independent sums of range $N$ after applying \eqref{U(N)Integral}. There are several types of independent sums which can all be tracked diagrammatically.
\begin{itemize}
\item Each cycle consisting of alternating dashed edges and edges of color 1 (resp. 3) corresponds to a sum over an index with the color 1 (resp. 3), hence it brings a factor $N$. The number of such cycles is denoted $F_{1}$ (resp. $F_{3}$).
\item Each cycle consisting of alternating dashed edges and edges with a box corresponds to a sum over indices with the colors 2 and 4, hence it brings a factor $N^2$. The number of such cycles is denoted $F_{\square}$.
\item The Weingarten function $\Wg_{N^2}(\sigma\tau^{-1})$ is a function of the cycle structure of $\sigma\tau^{-1}$, which we write $(1^{p_1} 2^{p_2} \dotsb n^{p_n})$ (meaning $p_j$ cycles of length $j$, with $\sum_j j p_j = n$). Each cycle is represented in the diagram as a cycle consisting of alternating dashed edges and edges of $U$ and $\overline{U}$. The number of cycles is denoted $F_0 = \sum_j p_j$. Moreover, the Weingarten function asymptotically behaves as
\begin{equation}
\Wg_{N^2}(1^{p_1} 2^{p_2} \dotsb n^{p_n}) \underset{\text{large $N$}}{=} (N^2)^{F_0-2n} \prod_{j=1}^n \Bigl[(-1)^{j-1}\, \operatorname{Cat}_{j-1}\Bigr]^{p_j}.
\end{equation}
That implies that each cycle comes with a factor $N^2$.
\end{itemize}
We can now evaluate the variation of all those quantities between the left and the right of figure \ref{fig:LeafContraction}.
\begin{itemize}
\item The two dashed lines incident to the black vertices on the left are parts of either one or two cycles with color 1 (depending on the details inside the rectangle) which gives two or one cycles on the right, hence $|\delta F_1|\leq 1$.
\item The two dashed lines incident to the black vertices on the left belong to one cycle with color 3, which splits into two cycles on the right, $\delta F_3 =1$.
\item The two dashed lines incident to the white vertices on the left belong to one cycle of type ``box'', which splits into two cycles on the right, $\delta F_{\square} =1$.
\item The four dashed lines on the left can either belong to two different cycles of the permutation $\sigma\tau^{-1}$ (one going along the edge $(\bar{v}_1 v_1)$ and the other along $(\bar{v}_2 v_2)$) or to the same one. On the right, we have a cycle $(\bar{v}_1 v_1 v_2 \bar{v}_2)$ and at least another one in the rectangle. Thus, $\delta F_0 \geq 0$.
\end{itemize}
Therefore the amplitude for the diagram on the right bounds the one for the diagram on the left by at least a factor $N^{2\delta F_{\square}} = N^2$. This means that for every configuration like the left hand side of figure \ref{fig:LeafContraction}, there is an associated configuration which is enhanced. The same conclusion is reached by means of similar arguments if instead of the left hand side of figure \ref{fig:LeafContraction} there is a dashed edge between $v_1$ and $v_2$ or between $\bar{v}_1$ and $\bar{v}_2$.

As a conclusion, it comes that all dominant contributions to the angular integral have a dashed edge between $v_1$ and $v_2$ and one between $\bar{v}_1$ and $\bar{v}_2$, which fixes one value of $\sigma$ and one value of $\tau$. It forces $\sigma\tau^{-1}$ to have a fixed point and this cycle contributes to a trivial factor 1 of the Weingarten function. Therefore, up to factors of $N$,
\begin{equation}
\begin{aligned}
\int_{U(N^2)} B_{\mathcal{T}}(U,\{\lambda_i\}) dU &= \int_{U(N^2)} dU \sum U_{\begin{smallmatrix} a_1\\ a_3\end{smallmatrix} \begin{smallmatrix} \alpha_2\\ \alpha_4\end{smallmatrix}} \lambda_{\begin{smallmatrix} \alpha_2\\ \alpha_4\end{smallmatrix}}^{2l} \overline{U}_{\begin{smallmatrix} b_1\\ a_3\end{smallmatrix}\begin{smallmatrix} \alpha_2\\ \alpha_4\end{smallmatrix}}\,\Bigl[f(UD^2U^\dagger)\Bigr]_{b_1 a_1} \\
&\underset{\text{large $N$}}{=} \Bigl(\sum_{\alpha_2, \alpha_4} \lambda_{\begin{smallmatrix} \alpha_2\\ \alpha_4\end{smallmatrix}}^{2l}\Bigr) \int_{U(N^2)} dU\,\sum_{a_1} \Bigl[f(UD^2U^\dagger)\Bigr]_{a_1 a_1}.
\end{aligned}
\end{equation}
The expectation of this product factorizes at large $N$, so that $\sum_{\alpha_2, \alpha_4} \lambda_{\begin{smallmatrix} \alpha_2\\ \alpha_4\end{smallmatrix}}^{2l} = N^2 \operatorname{Cat}_{l}$ and the trace of $f(MM^\dagger)$ is simply the bubble polynomial labeled by the tree $\mathcal{T}\setminus (\text{leaf, incident edge})$, where the two corners incident to the edge on the parent vertex are merged and their labels added. An induction from the leaves to the root leads to \eqref{CatProd}.

\section*{Conclusion}

In this short note, both approaches of \cite{MeandersTensors} and \cite{FullyPacked} have been pursued:
\begin{itemize}
\item exploring the relations between tensor and matrix models like in \cite{FullyPacked} and using them 
\item to evaluate new Gaussian expectations in random tensor theory, in particular for non-melonic polynomials, supplementing the results of \cite{MeandersTensors} through another method.
\end{itemize}

The starting point is as simple as that of \cite{FullyPacked}: a tensor can be seen as a (typically rectangular) matrix with one index being a $p$-uple of tensor indices and the other index a $(d-p)$-uple ($d$ being the total number of indices). Just as in \cite{FullyPacked}, unitary transformations and symmetries play a major role. But instead of offering a novel presentation of the results of random tensor theory, we rather apply tools from random matrices, namely integrals over the unitary group.


In the section \ref{sec:Gaussian}, we have embedded our tensor on a matrix space, but the polynomials of interest in tensor theory do not have a group of unitary symmetries as large as the matrix trace-invariants. As a consequence, a typical tensor observable depends on the angular degrees of freedom in addition to the singular values of the matrix. In a Gaussian distribution, integrating the observable over its angular variables leads to a notion of effective observable which writes as a sum of products of traces. This approach enables exact calculations in concrete cases as shown in section \ref{sec:ExactCalculation}.

We have shown in section \ref{sec:CatalanTrees} that in more complicated cases (i.e. when the degree of the polynomial to be integrated over the unitary group can be arbitrarily large) asymptotic calculations may still be possible. This was illustrated on a family of observables built from matrix trace-invariants glued together in a tree-like fashion.

As well-known, the large $N$ limit of tensor models (equipped with their standard scaling) with melonic interactions is Gaussian \cite{universality}: from a matrix viewpoint, it is as if the singular values all localize at the minimum of the potential \cite{toy-doublescaling,SDEs,IntermediateT4}. The method used here in \ref{sec:CatalanTrees} allows to evaluate expectations for \emph{non-melonic} observables, ones for which the distribution of singular values has a non-vanishing width. This is a new addition to the toolbox for random Gaussian tensors, complementing the meander approach of \cite{MeandersTensors} which fits a different set of observables. 

An open challenge is to take the exponential of such observables, so as to get non-melonic, non-Gaussian measures, which as shown in \cite{new1/N} do have non-Gaussian large $N$ limit (but whose entropy exponents are still unknown). Even the simplest integral over $U(N)$ in the Gaussian case was shown to be of degree 4, implying that the exponential is not of the Itzykson-Zuber type. It may be that techniques not considered in the present paper work, like the one developed in \cite{U(N)IntegralsEynard}, and this will be investigated elsewhere. We hope that the relationship we have established between tensor models and loop models can lead to fruitful cross-fertilization.

We would like to eventually mention a more speculatively related idea. The same way the Schwinger-Dyson equations of ordinary matrix models constitute a set of Virasoro constraints, those of tensor models can be cast as a set of constraints generated by operators which form a Lie algebra \cite{tree-algebra, bubble-algebra}. Those operators are differential with respect to the coupling constants and are labeled by bubbles. They basically trade the expectations of bubbles for linear operations. In order to understand this algebra, it seems natural to gain knowledge of those expectations (methods as well as exact or approximated evaluations). Only the large $N$ limit and the double scaling limit are so far understood through Schwinger-Dyson equations (they however only include melonic contributions and specific chains) \cite{SDEs, DSSD}. The calculation of Gaussian expectations for more generic bubbles is a first step in that direction. This is certainly crucial to quantum gravity or rather ``bubble theory'', as well as to the possible integrable properties of tensor models.

\section*{Acknowledgements}

Research at Perimeter Institute is supported by the Government of Canada through Industry Canada and by the Province of Ontario through the Ministry of Research and Innovation.


\end{document}